\documentclass{webofc}

\usepackage[varg]{txfonts}   
\usepackage{hyperref}
\usepackage{url}
\hypersetup{colorlinks=true,citecolor=blue,urlcolor=blue,linkcolor=blue}
\newcommand{\mee}{m_\mathrm{ee}}

\newcommand{\GeV}{\mathrm{GeV}}
\newcommand{\MeV}{\mathrm{MeV}}
\newcommand{\sqrtsnn}{\sqrt{s_\mathrm{NN}}}
\newcommand{\avg}[1]{\left\langle#1\right\rangle}
\usepackage{physics}
\begin{document}
\title{Imprints of dynamic ﬂuidization on dilepton production}
%
%

\author{\firstname{Renan} \lastname{G\'oes-Hirayama}\inst{1,2,3}\fnsep\thanks{\email{hirayama@itp.uni-frankfurt.de}} \and
        \firstname{Zuzana} \lastname{Paul\'inyová}\inst{4} \and
        \firstname{Joscha} \lastname{Egger}\inst{3} \and
        \firstname{Iurii} \lastname{Karpenko}\inst{5} \and\\
        \firstname{Hannah} \lastname{Elfner}\inst{6,1,2,3}
}

\institute{Helmholtz Forschungsakademie Hessen für FAIR (HFHF), Germany
\and
           Frankfurt Institute for Advanced Studies, Germany
\and
           Institut für Theoretische Physik, Goethe Universität, Germany
\and
           Faculty of Science, P.J. Šafárik University, Slovakia
\and
           Faculty of Nuclear Sciences and Physical Engineering, Czech Technical University in Prague
\and
			GSI Helmholtzzentrum für Schwerionenforschung, Germany
          }

\abstract{

We present a newly developed hybrid hadronic transport + hydrodynamics framework geared towards heavy ion collisions (HICs) at low to intermediate beam energies, and report on the resulting excitation function of dileptons. In this range of energies, it is unclear how to properly initialize the hydrodynamic evolution. Due to the cumulative electromagnetic radiation throughout the collision, dilepton observables are sensitive to the initial condition. In this work, we study how the dilepton ``thermometer'' is affected by employing dynamical initial conditions, in contrast to the traditional fixed-time approach.}
\maketitle

\section{Dilepton radiation in a hybrid approach}\label{sec:dynflu:dilepton}


Dileptons are frequently dubbed as multimessengers of heavy-ion collisions, with a prominent application being that of a ``thermometer'' \cite{Rapp:2014hha}. The electromagnetic radiation depends on the thermodynamic conditions of the system whence it is produced. The radiation from a thermal source with temperature $T$ has the shape
\begin{equation}\label{eq:thermal}
\dv{N}{\mee}\propto \mee^{3/2}\exp\left(-\frac{\mee}{T}\right),
\end{equation}
that is, it increases strongly with the local temperature, so the typical dilepton in the intermediate mass range (IMR) originates early on in the evolution. If the dilepton spectra can be associated to an exponential shape like \eqref{eq:thermal}, the logarithmic slope of the curve will correspond to an effective temperature close to the maximum temperature attained by the system. This implies that, in the context of an evolution model, they are particularly sensitive to how the initial state is constructed.

In \cite{Churchill:2023zkk}, thermal rates calculated with perturbative QCD were used to radiate IMR dielectrons from a \texttt{MUSIC-iS3D-UrQMD} hybrid approach, where the hydrodynamic evolution starts at a constant $\tau$, namely the \emph{nuclear passing time} $\tau_0=2R/\gamma_L v_L$, where $\gamma_L = \sqrtsnn/(2m_N)$ is the longitudinal Lorentz factor with $m_N = 0.938\ \GeV$, and $R$ is the radius of the colliding nuclei \cite{Denicol:2018wdp}. In this scenario, the maximum temperature is simply the initial value. A remarkable correlation was found between this value and the slope parameter extracted from the dilepton spectra, for Au+Au collisions at several energies and centrality classes.

This work is based on the \texttt{SMASH-vHLLE} hybrid model \cite{Schafer:2021csj, Gotz:2025wnv}. It is comprised of \texttt{SMASH}, a hadronic transport approach that serves  both as a framework for the initial conditions and for the late stage dynamics, and \texttt{vHLLE}, a solver for viscous hydrodynamics based on the Israel-Stewart equations \cite{Karpenko:2013wva}. Hybrid models have had great success in reproducing experimental data at intermediate and high beam energies.

\begin{figure}[h]
\centering
\includegraphics[width=0.48\textwidth]{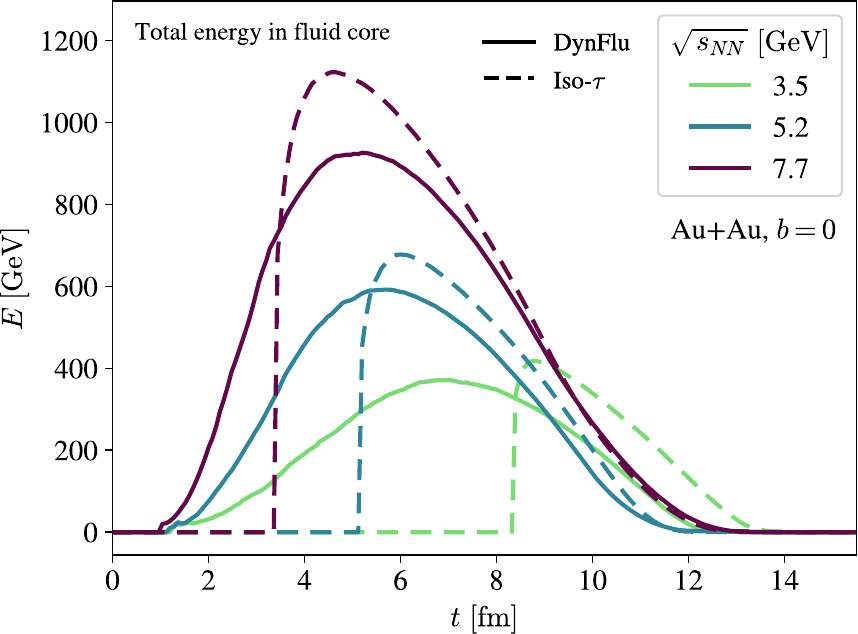}%
\hspace{0.015\textwidth}%
\includegraphics[width=0.48\textwidth]{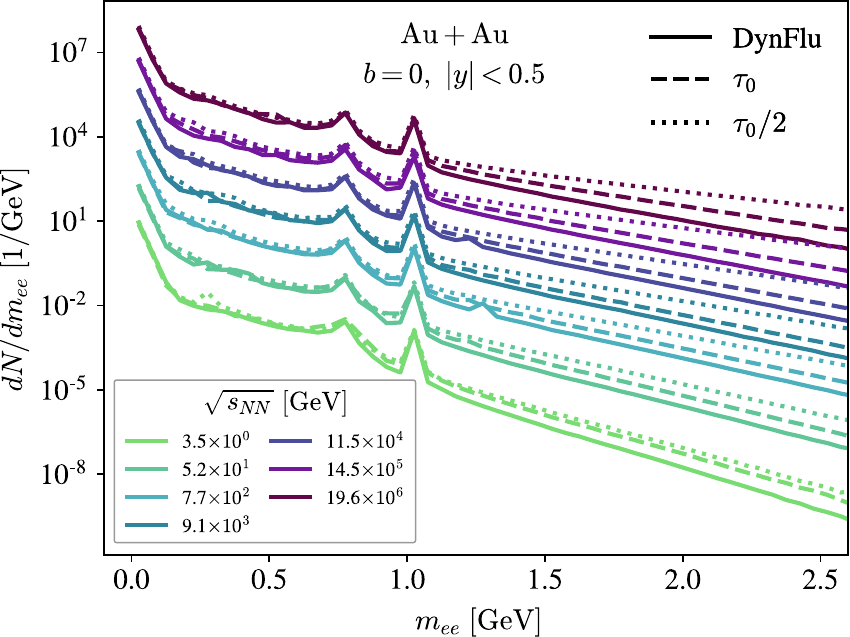}
\caption{Effect of different fluidization conditions for intermediate beam energies. Left: Total energy deposition in fluid core. Right: Invariant mass spectra of dileptons.}\label{fig:energy_spectra}
\end{figure}

Recently, we introduced dynamical initial conditions to this hybrid model
based on the local conditions surrounding a hadron in the initial hadronic evolution. If the energy density is larger than a predefined threshold, the hadron is used as a source for the hydrodynamic equations \cite{Goes-Hirayama:2025nls}. We call this approach \emph{dynamic fluidization} (DynFlu). At low beam energies $(\sqrtsnn<10\ \GeV)$, it is favorable in contrast to the traditional \emph{iso-$\tau$} initialization: since $\gamma_L$ is small, the nuclei are not flat disks and the constant $\tau_0$ becomes large, thus some regions of the fireball approach hydrodynamization sooner than others. As shown in the left plot of figure~\ref{fig:energy_spectra}, the energy is deposited in the fluid much more gradually within DynFlu than in iso-$\tau$. Besides that, the selection via energy density provides a natural core-corona separation, which is important as not all of the system is expected to fluidize. 
 
In this work, we study the emission of dielectrons at low-to-intermediate energies using the \texttt{SMASH-vHLLE} hybrid. The off-equilibrium contribution is calculated in \texttt{SMASH} via the \emph{shining method} \cite{Heinz:1991fn,Li:1994cj}, where each dilepton source radiates perturbatively at every time step. This provides a good description of dielectrons in elementary collisions \cite{Staudenmaier:2017vtq}, but with insufficient broadening of spectral functions of vector mesons in heavy-ion collisions \cite{Hirayama:2022rur}, and has dynamics consistent with flow measurements at HADES \cite{Goes-Hirayama:2024aqz}. The thermal radiation is calculated from the thermodynamic information in \texttt{vHLLE}, using the rates calculated within the Rapp-Wambach-Hees model \cite{Rapp:1999us,vanHees:2007th,Rapp:2014hha}, including both hadronic and QGP contributions.

The right plot of figure~\ref{fig:energy_spectra} shows the total spectra in the range $3.5\leq\sqrtsnn/\GeV\leq19.6$ calculated within our hybrid using both DynFlu and iso-$\tau$ initial conditions. We also compare it to the limiting case of fluidizing the hadrons at the maximum overlap time $\tau_0/2$, the earliest time when every nucleon can interact, as a lower bound, and to infer the sensitivity on a specific initialization time.

In the low mass range ($\mee\leq1.2\ \GeV$), the dilepton yields are not significantly impacted by changing the fluidization condition, which reflects that the core-corona separation is done appropriately. The ratios between the models increase noticeably in the IMR, as expected from the left plot: the bigger differences take place at the beginning of the hydrodynamic evolution, when the medium is hot. The IMR dileptons are emitted primarily from the hot and dense stages, so the initialization time plays a decisive role. In addition, the effect of the specific initialization time grows with beam energy.

\section{Dielectrons as a thermometer}

\begin{figure}[ht]
\centering
\includegraphics[width=0.48\linewidth]{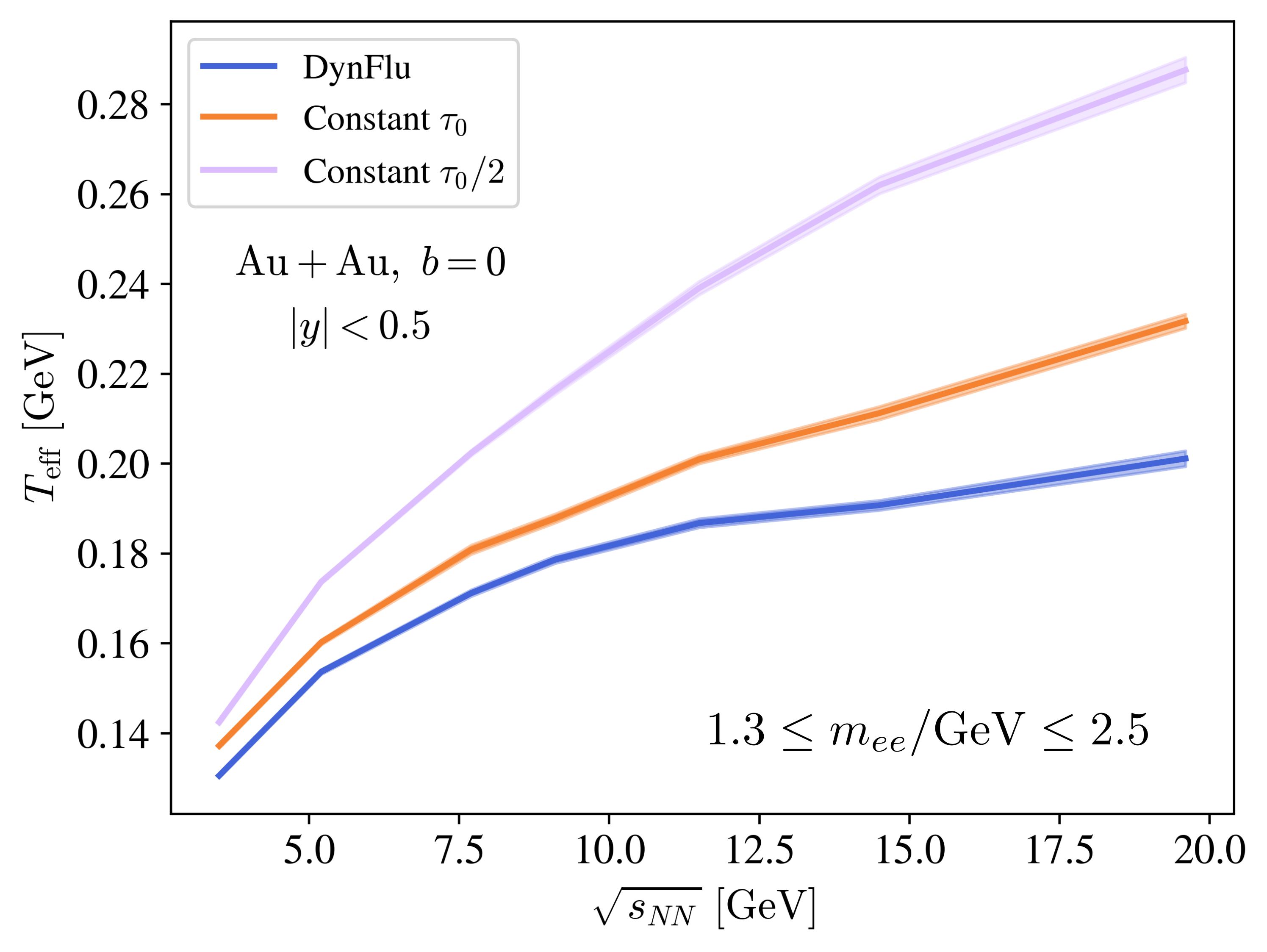}%
\hspace{0.01\textwidth}
\includegraphics[width=0.48\linewidth]{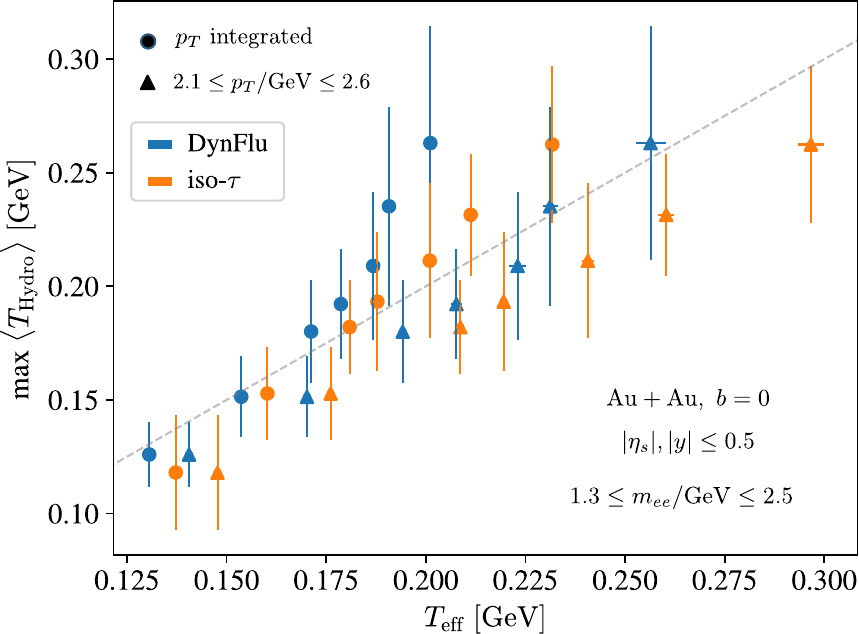}%
\caption{Left: beam energy dependence of the effective dilepton temperature for different fluidization setups. Right: correlation between said  temperature with the maximum average temperature reached by the fluid. The bars around each point indicate the fit uncertainty in $T_\mathrm{eff}$ and the $1\sigma$ spatial variation in $\avg{T_\mathrm{Hydro}}$.}\label{fig:excitation_correlation}
\end{figure}

We can determine an effective temperature for the IMR radiation from the slope parameter, by fitting the spectra in figure~\ref{fig:energy_spectra} to \eqref{eq:thermal} in the IMR, which we define as $1.3\leq\mee/\GeV\leq2.5$. The resulting excitation functions for the temperatures are shown in the left plot of figure~\ref{fig:excitation_correlation}.

The extracted temperatures within DynFlu and iso-$\tau$ (with $\tau_0$) differ from $\Delta T_\mathrm{eff}\approx8\ \MeV$ at $\sqrtsnn=3.5\ \GeV$ to $\Delta T_\mathrm{eff}\approx30\ \MeV$ at $\sqrtsnn=19.6\ \GeV$. In comparison to the $\tau_0/2$ curve, which serves as an upper bound, the differences are much greater, emphasizing the sensitivity of this extraction to the initial state.

The dilepton slope was proposed in \cite{Churchill:2023zkk} as a thermometer of the initial medium temperature by way of the linear fit
\begin{equation}\label{eq:correlate_T}
\avg{T_\mathrm{in}}=\max\limits\avg{T_\mathrm{Hydro}}=\kappa T_\mathrm{eff}+c,
\end{equation}
where $T_\mathrm{in}$ is the initial temperature of the fluid in an iso-$\tau$ framework, and the brackets denote an average weighted by energy density. This is equivalent to the maximum average temperature $\max\avg{T_\mathrm{Hydro}}$, which is a more meaningful quantity in DynFlu. Ref. \cite{Churchill:2023zkk} found $\kappa=1.55(2)$ and $c=-0.093(3)\ \GeV$, so indeed, $T_\mathrm{eff}$ is well-correlated with $T_\mathrm{Hydro}$. However, the former is quite sensitive to the fluidization approach while the latter is not \cite{Goes-Hirayama:2025nls}. The right plot of figure~\ref{fig:excitation_correlation} shows the correlation in our calculation. Within iso-$\tau$ (using $\tau_0$, orange circles) provides compatible values to Ref. \cite{Churchill:2023zkk}, as shown in table~\ref{tab:fit}, ensuring a faithful comparison. On the other hand, the DynFlu approach (blue circles) has a considerably larger slope. This discrepancy cannot be easily seen in experiment, and hinders the reading of a “dilepton thermometer”. 

\begin{table}
\centering
\caption{Fit parameters for \eqref{eq:correlate_T} extracted from figure~\ref{fig:excitation_correlation}. Values of $c$ are given in $\GeV$.}
\label{tab:fit}       
\begin{tabular}{c|cc|cc}
            & \multicolumn{2}{c|}{$p_T$ integrated} & \multicolumn{2}{l}{$2.1\leq p_T/\GeV\leq2.6$} \\ \cline{2-5} 
            & DynFlu & iso-$\tau$& DynFlu & iso-$\tau$ \\ \hline
$\kappa$       & $1.88$  & $1.52$   &  $1.19$    &  $0.95$  \\
$c$ &    $-0.13$    &   $-0.09$  & $-0.05$  &     $-0.02$   
\end{tabular}
\end{table}

The $p_T$ spectrum has the same exponential dependence on temperature as \eqref{eq:thermal} in thermal equilibrium, thus dileptons at higher transverse momentum are also correlated with an earlier emission. Performing the fit \eqref{eq:correlate_T} only for particles in a high $p_T$ (triangles on the right of figure~\ref{fig:excitation_correlation}) bin, we find fit parameters much closer to the identity $\max\avg{T_\mathrm{Hydro}}=T_\mathrm{eff}$, indicating an improved notion of thermometer.

\section{Conclusion}

We showed that the choice of initialization of hydrodynamics in a hybrid model affects the production of dileptons in the intermediate  range of invariant masses. With dynamic initial conditions, the deposition of energy is gradual, and consequently the effective temperature $T_\mathrm{eff}$ extracted from dilepton yields is lower. While $T_\mathrm{eff}$ correlates reasonably with the maximum temperatures achieved, the uncertainty in the initialization method makes this thermometer less reliable. Performing a $p_T$ cut provides 	a closer relation between the high medium temperatures and the dilepton spectra.

%
%
%
\bibliography{QM25_Hirayama_proc}
\end{document}